%% file: main.tex
\documentclass[sigconf, authorversion]{acmart}

\usepackage{gensymb}
\usepackage{enumitem}
\usepackage{subcaption}
\usepackage{graphicx}
\usepackage{changepage}

\usepackage[final]{changes}

\AtBeginDocument{%
  \providecommand\BibTeX{{%
    \normalfont B\kern-0.5em{\scshape i\kern-0.25em b}\kern-0.8em\TeX}}}

\copyrightyear{2022}
\acmYear{2022}
\setcopyright{acmcopyright}\acmConference[ASSETS '22]{The 24th International ACM SIGACCESS Conference on Computers and Accessibility}{October 23--26, 2022}{Athens, Greece}
\acmBooktitle{The 24th International ACM SIGACCESS Conference on Computers and Accessibility (ASSETS '22), October 23--26, 2022, Athens, Greece}
\acmPrice{15.00}
\acmDOI{10.1145/3517428.3544802}
\acmISBN{978-1-4503-9258-7/22/10}

\acmSubmissionID{9786}

\begin{document}

\title[Uncovering Visually Impaired Gamers’ Preferences for SATs Within Video Games]{Uncovering Visually Impaired Gamers’ Preferences for Spatial Awareness Tools Within Video Games}


\author{Vishnu Nair}
\affiliation{%
  \institution{Columbia University}
  \city{New York}
  \state{New York}
  \country{USA}
}

\author{Shao-en Ma}
\affiliation{%
  \institution{Columbia University}
  \city{New York}
  \state{New York}
  \country{USA}
}

\author{Ricardo E. Gonzalez Penuela}
\affiliation{%
  \institution{Cornell University, Cornell Tech}
  \city{New York}
  \state{New York}
  \country{USA}
}

\author{Yicheng He}
\affiliation{%
  \institution{Columbia University}
  \city{New York}
  \state{New York}
  \country{USA}
}

\author{Karen Lin}
\affiliation{%
  \institution{Columbia University}
  \city{New York}
  \state{New York}
  \country{USA}
}

\author{Mason Hayes}
\affiliation{%
  \institution{Rochester Institute of Technology}
  \city{Rochester}
  \state{New York}
  \country{USA}
}

\author{Hannah Huddleston}
\affiliation{%
  \institution{Stanford University}
  \city{Stanford}
  \state{California}
  \country{USA}
}

\author{Matthew Donnelly}
\affiliation{%
  \institution{Bowdoin College}
  \city{Brunswick}
  \state{Maine}
  \country{USA}
}

\author{Brian A. Smith}
\affiliation{%
  \institution{Columbia University}
  \city{New York}
  \state{New York}
  \country{USA}
}

\renewcommand{\shortauthors}{Nair et al.}

\begin{abstract}
  Sighted players gain spatial awareness within video games through sight and spatial awareness tools (SATs) such as minimaps. Visually impaired players (VIPs), however, must often rely heavily on SATs to gain spatial awareness, especially in complex environments where using rich ambient sound design alone may be insufficient. Researchers have developed many SATs for facilitating spatial awareness within VIPs. Yet this abundance disguises a gap in our understanding about how exactly these approaches assist VIPs in gaining spatial awareness and what their relative merits and limitations are. To address this, we investigate four leading approaches to facilitating spatial awareness for VIPs within a 3D video game context. Our findings uncover new insights into SATs for VIPs within video games, including that VIPs value position and orientation information the most from an SAT; that none of the approaches we investigated convey position and orientation effectively; and that VIPs highly value the ability to customize SATs.
\end{abstract}

\ccsdesc[500]{Human-centered computing~Auditory feedback}
\ccsdesc[300]{Human-centered computing~Accessibility technologies}
\ccsdesc[100]{Human-centered computing~Accessibility systems and tools}

\keywords{Audio navigation tools; spatial awareness tools; blind-accessible games; visual impairments}

\begin{teaserfigure}
  \centering
  \includegraphics[width=0.85\textwidth]{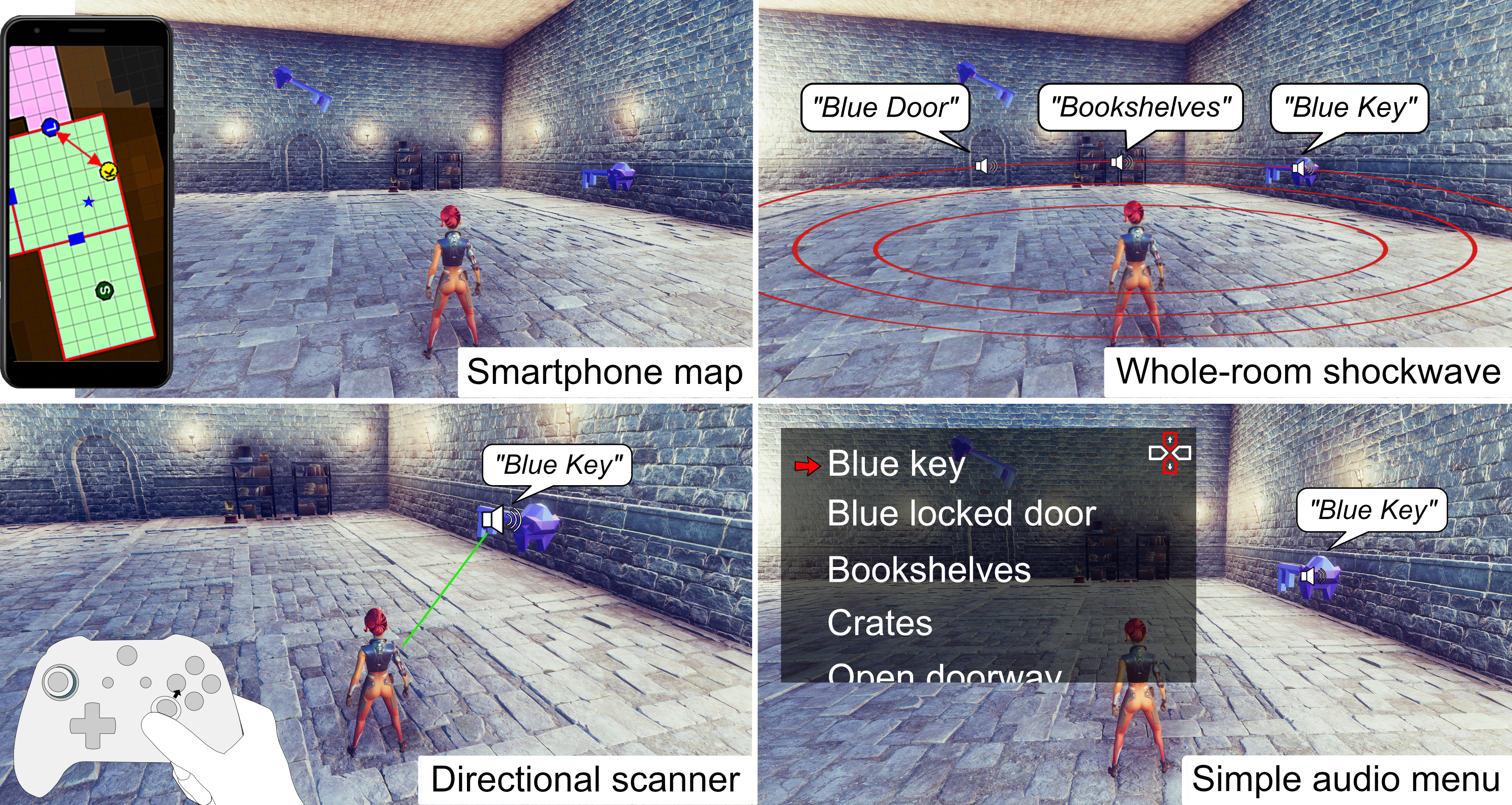}
  \caption{Illustrations of the four spatial awareness tools we implemented within \textit{Dungeon Escape}. These approaches --- the smartphone map, the whole-room shockwave, the directional scanner, and the simple audio menu --- represent a broad range of designs for facilitating spatial awareness for VIPs. However, we still do not yet understand what the relative merits and limitations of each approach are.}
  \Description{Four views of a video game character within a room. In each view, the player is using a different spatial awareness tool.}
  \label{fig:tools}
\end{teaserfigure}

\maketitle

\section{Introduction}
\input{sec01-intro}


\section{Background and Related Work}
\input{sec02-rw}


\section{Four Spatial Awareness Tools}
\label{section:tooldesc}
\input{sec03-fourtools}


\section{User Study}
\input{sec04-userstudy}


\section{RQ1 Results: Aspects of Spatial Awareness Important to Visually Impaired Players}
\label{sec:affresults}
\input{sec05-RQ1results}


\section{RQ2 Results: Comparison of Spatial Awareness Tools for Virtual Worlds}
\label{sec:toolresults}
\input{sec06-RQ2results}


\section{Discussion: Takeaways from RQ1 and RQ2 Together }
\label{sec:discussion_section}
\input{sec07-discussion}


\section{Future Work}
\label{sec:disc}
\input{sec08-futurework}


\section{Limitations}
\input{sec09-limitations}


\section{Conclusion}
\input{sec10-conclusion}


\begin{acks}
We would like to thank Michael Malcolm and Sebasti\'{a}n Mercado for their assistance during our pilot tests. We would also like to extend our sincere gratitude toward our study participants for their participation and to the anonymous reviewers for their helpful feedback. Mason Hayes, Hannah Huddleston, and Matthew Donnelly were funded by National Science Foundation Grants 2051053 and 2051060. The opinions, findings, conclusions, and/or recommendations expressed are those of the authors and do not necessarily reflect the views of the National Science Foundation.
\end{acks}

\bibliographystyle{ACM-Reference-Format}
\bibliography{refs}

\end{document}

%% file: sec01-intro.tex
Mainstream 3D video games are largely inaccessible to visually-impaired players (VIPs) because they often lack crucial accessibility tools~\cite{Archambault2008, Porter2013}. Although some recent mainstream games~\cite{NaughtyDog2020} have made strides in making certain in-game abilities accessible to VIPs, many crucial abilities remain inaccessible. Among these is the ability for players to gain spatial awareness of their surroundings, which prior work has established is crucial for granting VIPs with an enhanced sense of space and presence within the game world~\cite{Andrade2018, Andrade2021}.

Sighted players gain spatial awareness using a combination of vision and auxiliary tools such as minimaps and world maps~\cite{Bork2019, Cossu2019Ch10, Cossu2019Ch12, Thorn2018, Zagata2021}, which we will refer to as \textit{spatial awareness tools (SATs)}. VIPs, however, are not able to benefit from vision in the same way as sighted players. Although rich environmental sound design can ensure some level of accessibility and spatial awareness for VIPs, these elements can prove to be insufficient, especially in complex virtual environments. As such, SATs are often an indispensable view of the game world for VIPs, having an immense impact on their experience within a game.

Researchers have developed several types of SATs which represent very different approaches for facilitating a representation of the game world for VIPs. These include touchscreen maps, shockwave-like systems, directional scanners, and audio-based menus. This relative abundance, however, disguises a gap in our understanding about how exactly these approaches assist VIPs in gaining spatial awareness and what the relative merits and limitations of each approach are. It is not yet clear to developers, for example, whether it is best to use a shockwave-like system like \textit{The Last of Us Part 2} did~\cite{NaughtyDog2020} or audio-based menus like \textit{Terraformers} did~\cite{Westin2004, PinInteractive2003a}, and blind gamers are ultimately the ones who suffer. As such, we take a step back and ask two important research questions:

\begin{enumerate}[label=\textbf{RQ\arabic*:}, leftmargin=3.5\parindent]
    \item What aspects of spatial awareness do VIPs find important \textit{within games}?
    \item How well do today's differing SAT approaches facilitate each aspect of spatial awareness, and why?
\end{enumerate}

In this work, we investigate RQ1 and RQ2 by implementing four leading approaches to facilitating spatial awareness for VIPs and investigating their merits and limitations. The four tools, illustrated in Figure \ref{fig:tools}, are a smartphone map, a whole-room shockwave, a directional scanner, and a simple audio menu of points-of-interest. Together, they represent a broad range of design choices, including touchscreen-based vs. game controller interaction and "all-at-once" (collective) overviews vs. pointer-based scanning.

In order to investigate these research questions, we conducted a user study in which nine visually impaired participants played multiple levels of a 3D adventure video game using our tools. 

For RQ1, we evaluated how important VIPs consider six different types of spatial awareness --- that have been found to be important in physical world settings~\cite{RowellUngar2005, GiudiceLegge2008, Giudice2020, Kacorri2016, Hill1993, Yatani2012, Klatzky1998, Epstein2017} --- to be in comparison to each other within a video game context. Section ~\ref{sec:affdesc} identifies the six types. We observed that participants considered position and orientation to be the most important type of spatial awareness and that they considered the scale and shape of an area to be the least important aspects of spatial awareness. These particular findings reflect existing work within the physical world that highlights the importance of position and orientation for VIPs~\cite{RowellUngar2005, Kacorri2016, GiudiceLegge2008}, but differ from existing work within the physical world that has also found knowledge of area scale and shape to be an important component of better understanding an area, especially when freely exploring it~\cite{HolmesArditi1998, Andrade2021}.

With respect to RQ2, we observed that each tool had its own strength for VIPs: The directional scanner communicated the arrangement of items very well; the simple audio menu communicated the presence of items very well; the smartphone map communicated the shape of an area very well; and the whole-room shockwave communicated the scale of the area well. Importantly, however, we also discovered significant deficiencies in today's spatial awareness tools. No tool excelled across the board, and in particular, none of the tools communicated position and orientation very well despite our finding from RQ1 that position and orientation is the most important type of spatial awareness to VIPs within games. Furthermore, we found issues with the tools that influenced how effectively they communicated spatial awareness to players, including that some of the tools provided too much information. 

Together, our findings from RQ1 and RQ2 reveal important design implications for future spatial awareness tools for VIPs within video games, and we present these in our Discussion. We also discuss the potential for developing purpose-built hardware for spatial awareness and how our findings within virtual worlds can inspire further research in physical world navigation and exploration. 

%% file: sec02-rw.tex
Our work is built upon a rich history of prior work on facilitating spatial awareness for VIPs, both within the physical world and within video games.

We begin this section by explaining what we mean by ``spatial awareness'' --- in particular, by reviewing aspects of spatial awareness that are known to be important to VIPs within physical world contexts (Section~\ref{sec:affdesc}). RQ1 will investigate the relative importance of these aspects for VIPs \textit{within games}. We then review existing techniques for facilitating spatial awareness for VIPs, both in the physical world and within video game environments --- and review the tradeoffs inherent within the design of these tools (Section~\ref{sec:rw-sub2}). Through RQ2, we take a step back and investigate the relative merits and limitations of these approaches.

\subsection{What do we mean by ``spatial awareness''?}
\label{sec:affdesc}

Spatial awareness, as used in this work, refers to a user's awareness of their surrounding environment and of their own state within the environment~\cite{Klippel2010, Yang2011}. Past literature within physical world contexts has shown spatial awareness to be multifaceted. Thus, in this work, we investigate RQ1 and RQ2 with respect to \textit{six} distinct aspects of spatial awareness we identified through prior work. Specifically, we looked through existing research in cognitive map formation and spatial awareness for VIPs within the physical world and looked for explicit information on what aspects of spatial awareness are most important to VIPs. We chose to investigate the following six types of spatial awareness since they were mentioned as important across a breadth of prior research~\cite{RowellUngar2005, GiudiceLegge2008, Giudice2020, Kacorri2016, Hill1993, Yatani2012, Klatzky1998, Epstein2017}:\\


\begin{adjustwidth}{0.5cm}{}
    \noindent \textbf{Types 1 \& 2:} \textbf{\textit{Scale} and \textit{shape} of the area.} Prior work --- mainly in tactile maps~\cite{RowellUngar2005, HolmesArditi1998} and echolocation~\cite{Andrade2021} --- has found area shape to be important to VIPs in obtaining a general impression of the area, which can be especially crucial when exploring and trying to learn about the environment.\\ 
    
    \noindent \textbf{Type 3:} \textbf{\textit{Position \& orientation}.} Researchers have found that understanding where one is within a mental map of the area (for example, their Cartesian coordinates or their heading direction in degrees) --- that is, within an \textit{allocentric}~\cite{Klatzky1998}, map-like mental representation of the environment --- is vital to continuously updating their own current state within the environment and thus effectively move through it~\cite{Klatzky1998, Epstein2017, Giudice2020}. Yet, prior work in physical world contexts~\cite{GiudiceLegge2008} has shown that obtaining this understanding is especially demanding for VIPs.\\

    \noindent \textbf{Types 4 \& 5:} \textbf{\textit{Presence} and \textit{arrangement} of items.} Researchers have emphasized that providing VIPs with the information necessary to perceive the locations of objects can allow them to infer spatial relationships between objects and can lead to increased spatial awareness and more accurate cognitive maps~\cite{GiudiceLegge2008, Hill1993}.\\

    \noindent \textbf{Type 6:} \textbf{Areas \textit{adjacent} to the player's current area.} Prior work with physical world tactile maps~\cite{RowellUngar2005} and mobile-based spatial tactile feedback for communicating geographical information~\cite{Yatani2012} have underscored the importance of understanding the global structure of the world --- general overviews of an area and spatial relationships between multiple areas --- for VIPs, which can help them plan out routes and backtrack as needed.\\ 
\end{adjustwidth}

Although prior work has determined these six aspects of spatial awareness to be important to VIPs in the physical world, video games are very different from the physical world. Within the physical world, practicality and physical safety are extremely important factors~\cite{Banovic2013}, while in video games, agency and pleasure (fun) are very important, and VIPs' in-game ``safety'' may not always be a major concern. It is possible that, due to these differences, VIPs may find certain aspects of spatial awareness more or less important within games when compared to the physical world. We, thus, use RQ1 to explore these preferences.

\subsection{How do games supplement spatial awareness?}
\label{sec:rw-sub2}

Games made for VIPs often use ambient signals to provide \textit{implicit} spatial awareness to players. These ambient signals usually take the form of environmental audio cues that communicate information about the player’s immediate environment. For example, hearing running water may indicate that there is a waterfall or stream near the player. When environmental sounds reverberate, the player may realize that they are inside a cave or tunnel, and the extent of the reverberation can indicate the size of the cave or tunnel. The use of 3D sound can additionally communicate the relative direction that the source of sound is in with respect to the player.

Although ambient signals may be sufficient for simple environments, they can become less useful to players as environments become more complex, as is typical for many mainstream 3D games. Ambient cues can become overwhelming when there are too many items in the environment, and they may also be vague, giving players little information about what the sounds they are hearing actually represent. As a result, using ambient signals alone as a means to facilitate spatial awareness for players limits the complexity of games that accessible game designers are able to make. Accessible game designers, thus, face a tradeoff between designing environments that are interesting and designing games that are still accessible and playable by VIPs~\cite{Andrade2018, SmithNayar2018}.

Given the limitations of implicit forms of spatial awareness, accessible game designers often turn to creating tools that \textit{explicitly} communicate spatial awareness information to players. These spatial awareness tools (or SATs) --- which include (but are not limited to) tactile maps, radar systems, and grid systems --- supplement implicit spatial awareness cues by clarifying environmental information and affording players greater control over what information they hear and when they hear it.

Table~\ref{tab:sat-table} shows an overview of SATs from prior work. Below we review some explicit approaches for facilitating spatial awareness in games and in the physical world.


\input{sec02-TABLE-explicitSATs}


\subsubsection{Facilitating spatial awareness for VIPs within video games.}
\label{sec:rw-sat-virtual}

\hfill\newline Tools that explicitly communicate spatial awareness information to VIPs are not commonplace within mainstream video games.
Most examples, instead, come from ``audio games'' (audio-based games created for VIPs), which generally provide players with spatial awareness by presenting environments in the form of lists and grids that players can query. This technique is employed by many well-known audiogames, including \textit{Terraformers}~\cite{PinInteractive2003a, Westin2004}, \textit{A Hero’s Call}~\cite{OutOfSightGames2017}, and \textit{ShadowRine}~\cite{Matsuo2016}. These representations may communicate the presence and arrangement (Types 4~\&~5 from Section \ref{sec:affdesc}) of items and points-of-interest  and are sometimes further supplemented by additional tools such as radars and compasses.

Several examples of SATs have come from the research community as well. A notable example is NavStick~\cite{Nair2021pp, NairSmith2020}, which repurposes a game controller's right thumbstick to allow VIPs to ``look around'' their in-game surroundings via line-of-sight. A directional scanning system like NavStick could allow VIPs to determine the presence and spatial arrangement of objects around them (Types 4~\&~5) as well as their relative position and orientation within the game world (Type~3). 

A notable exception to the lack of SATs in mainstream games is \textit{The Last Of Us Part 2}, a 3D action-adventure game released in 2020~\cite{NaughtyDog2020}, which introduced an "enhanced listen mode" for VIPs. The enhanced listen mode provides spatial awareness to players by placing 3D audio beacons at the locations of nearby enemies and other points-of-interest on the press of a button. The beacons may give players a sense of the spatial arrangement of items in the area (Type~5) as well as a sense of the surrounding area's scale (Type~1).

\subsubsection{Facilitating spatial awareness for VIPs in the physical world.}
\label{sec:rw-sat-physical}

\hfill\newline Some audio-based tools within the physical world have features that explicitly provide VIPs with spatial awareness information and can thus inform the design of SATs for game worlds. NavCog3~\cite{Sato2017}, a turn-by-turn indoor navigation system for VIPs, for example, emits notifications about nearby landmarks and points-of-interest to promote awareness in the user of their presence (Type~4). Similarly, Microsoft Soundscape~\cite{MicrosoftResearch2018}, an audio-based wayfinding system that can be used by VIPs, uses 3D sound to communicate the presence and relative direction (i.e., arrangement, Type~5) of nearby landmarks. The spatial awareness that these systems provide is, however, limited. For example, they do not provide any information about the area's shape and size (Types 1~\&~2).

Tactile-based systems provide spatial awareness by providing overviews of areas~\cite{RowellUngar2005, HolmesArditi1998}, which may include the scale and shape of an area (Types 1~\&~2), the presence and arrangement of landmarks and other points-of-interest (Types 4~\&~5), and even what areas may be adjacent to a given area (Type~6). These not only include physical tactile maps but also mobile-based tactile systems, such as Timbremap~\cite{Su2010} and SmartTactMaps~\cite{Gotzelmann2015}, which can allow VIPs to survey the area they are in using a commodity smartphone.

Echolocation, which has been explored for both physical~\cite{Kish2009, ThalerGoodale2016, Norman2021} and virtual~\cite{Andrade2018, Andrade2020} environments, is another technique that VIPs may use to gain spatial awareness within environments. Using the acoustic properties of the environment can allow individuals to learn about the structure of the area they are in, including the scale and shape of the area (Types 1~\&~2), as well as the presence and arrangement of nearby objects (Types 4~\&~5)~\cite{ThalerGoodale2016, Kish2009}. 

%% file: sec02-TABLE-explicitSATs.tex
\begin{table*}[]
\centering
\resizebox{\textwidth}{!}{%
\begin{tabular}{l|l|l|l}
\textbf{Artifact} &
  \textbf{Ambient spatial awareness cues} &
  \textbf{Explicit spatial awareness tools (SATs)} &
  \textbf{SAT(s) in our study} \\ \hline \hline
\begin{tabular}[c]{@{}l@{}}\textit{ShadowRine}\\ \textit{(virtual)} \cite{Matsuo2016}\end{tabular} &
  \begin{tabular}[c]{@{}l@{}}Audio cues in environment.\\ (e.g., enemy \& object sounds)\end{tabular} &
  Tactile display showing top-down view of player's location. &
  Smartphone map \\ \hline
\begin{tabular}[c]{@{}l@{}}\textit{The Last of Us Part 2}\\ \textit{(virtual)} \cite{NaughtyDog2020}\end{tabular} &
  \begin{tabular}[c]{@{}l@{}}Audio cues in environment.\\ (with “audio cue glossary”)\end{tabular} &
  "Enhanced listen mode" (shockwave-like tool). &
  Whole-room shockwave \\ \hline
\begin{tabular}[c]{@{}l@{}}NavStick\\ \textit{(virtual)} \cite{NairSmith2020, Nair2021pp}\end{tabular} &
  \begin{tabular}[c]{@{}l@{}}Audio cues in environment.\\ (e.g., enemy \& checkpoint sounds)\end{tabular} &
  "NavStick" (direction-based object scanner). &
  Directional scanner \\ \hline
\begin{tabular}[c]{@{}l@{}}\textit{A Hero’s Call}\\ \textit{(virtual)} \cite{OutOfSightGames2017}\end{tabular} &
  Audio cues in environment. &
  Menu of nearby points of interest. &
  Simple audio menu \\ \hline
\begin{tabular}[c]{@{}l@{}}\textit{Swamp}\\ \textit{(virtual)} \cite{Kaldobsky2011}\end{tabular} &
  \begin{tabular}[c]{@{}l@{}}Audio cues in environment.\\ (e.g., zombie growls)\end{tabular} &
  \begin{tabular}[c]{@{}l@{}}“Radar” (beeps based upon empty space or solid walls).\\ Audio menu (with compass- \& tile-based guidance).\end{tabular} &
  \begin{tabular}[c]{@{}l@{}}Whole-room shockwave\\ Simple audio menu\end{tabular} \\ \hline
\begin{tabular}[c]{@{}l@{}}\textit{Terraformers}\\ \textit{(virtual)} \cite{Westin2004, PinInteractive2003a}\end{tabular} &
  Audio cues in environment. &
  \begin{tabular}[c]{@{}l@{}}“Sonar” (provides distance to object in current facing direction).\\ “GPS” (audio-based menu of nearby objects \& positions).\end{tabular} &
  \begin{tabular}[c]{@{}l@{}}Directional scanner\\ Simple audio menu\end{tabular} \\ \hline \hline
\begin{tabular}[c]{@{}l@{}}SmartTactMaps\\ \textit{(physical)} \cite{Gotzelmann2015}\end{tabular} &
  Sounds from physical environment. &
  Smartphone-based augmentation of physical tactile map. &
  Smartphone map \\ \hline
\begin{tabular}[c]{@{}l@{}}Timbremap\\ \textit{(physical)} \cite{Su2010}\end{tabular} &
  Sounds from physical environment. &
  Touchscreen-based 2D map exploration. &
  Smartphone map \\ \hline
\begin{tabular}[c]{@{}l@{}}Echolocation\\ \textit{(physical)} \cite{Kish2009, ThalerGoodale2016, Norman2021}\end{tabular} &
  Sounds from physical environment. &
  Behavior of reflected sounds within the environment. &
  Whole-room shockwave \\ \hline
\begin{tabular}[c]{@{}l@{}}Talking Points 3\\ \textit{(physical)} \cite{Yang2011}\end{tabular} &
  Sounds from physical environment. &
  “Directional Finder” (direction-based landmark scanner). &
  Directional scanner \\ \hline
\begin{tabular}[c]{@{}l@{}}MS Soundscape\\ \textit{(physical)} \cite{MicrosoftResearch2018}\end{tabular} &
  Sounds from physical environment. &
  \begin{tabular}[c]{@{}l@{}}Selection of points of interest from a menu.\\ Notifications about nearby landmarks using 3D sound.\end{tabular} &
  Simple audio menu \\ \hline
\begin{tabular}[c]{@{}l@{}}SpaceSense\\ \textit{(physical)} \cite{Yatani2012}\end{tabular} &
  Sounds from physical environment. &
  \begin{tabular}[c]{@{}l@{}}Vibration cues indicating the direction of a location \\ selected from a menu.\end{tabular} &
  Simple audio menu \\ \hline
\begin{tabular}[c]{@{}l@{}}NavCog3\\ \textit{(physical)} \cite{Sato2017}\end{tabular} &
  Sounds from physical environment. &
  Audio notifications of immediate surroundings. &
  N/A
\end{tabular}%
}
\vspace{0.5mm}
\caption{An overview of prior work in communicating spatial awareness to VIPs within both virtual and physical contexts. These artifacts represent a variety of ideas ranging from audio-based solutions to tactile solutions. For each, we present the ambient/implicit spatial awareness cues that it provides, the explicit spatial awareness tools (SATs) it introduces, and the corresponding SAT(s) in our study. The four SATs we implement collectively represent a significant portion of prior work.}
\label{tab:sat-table}
\end{table*}

%% file: sec03-fourtools.tex
In order to investigate RQ1 and RQ2, we implemented four existing approaches for giving VIPs spatial awareness of their surroundings that represent a wide range of possible designs. Figure \ref{fig:tools} depicts the four approaches. They include a smartphone map, a whole-room shockwave, a directional scanner, and a simple audio menu. We limited our exploration to just four tools to avoid fatiguing our user study participants while still effectively evaluating the tools. Regardless, Table~\ref{tab:sat-table} shows that these four tools collectively represent a significant portion of approaches from prior work on explicitly communicating spatial awareness information to VIPs. 

We were able to replicate two of the tools (the directional scanner and the simple audio menu) concretely from existing work~\cite{Nair2021pp, NairSmith2020, PinInteractive2003a, Westin2004}. For the smartphone map and the whole-room shockwave, however, we went through multiple design iterations because their implementation included many open design decisions that were not fully specified by prior work. In our design iterations, which we describe in Sections~\ref{sec:toolmap} and \ref{sec:toolshockwave}, our focus was to polish the tools’ designs and ensure that they showcased the potential of the two SAT approaches in the best possible way so that our results would not be confounded by a potentially bad design.

In order to ensure that our tools most accurately represented current approaches and to ensure that our study procedure was sound, we conducted pilot tests with two visually impaired and eight sighted-but-blindfolded people. We intended for the testing phase that included the sighted-but-blindfolded participants to be a na\"{i}ve-yet-useful way of catching any low-hanging fruit with respect to procedural, game-related, or tool-related issues before piloting with our visually impaired team members. Our visually impaired team members --- whom we hosted as part of the research team during the project --- then provided feedback that was critical to developing the final designs of the tools.

In the following subsections, we describe the design and implementation process of the four SATs. We direct readers to the accompanying video figure for a demonstration of all four tools. We created all four tools using the Unity game engine (v2020.3.16f1)~\cite{UnityTechnologies2020}.

\subsection{Smartphone Map}
\label{sec:toolmap}

The smartphone map interface, shown in the upper-left corner of Figure \ref{fig:tools}, uses a smartphone-based touchscreen map that works in tandem with the game. The player can use their finger to survey the map. As the player moves through the level, the map will automatically pan and rotate in real-time, allowing users to explicitly keep track of their own position and orientation, respectively.


The smartphone map interface represents prior work in tactile-based maps to support spatial awareness. Tactile maps in the physical world have been shown to support spatial awareness in VIPs by providing general overviews of spaces and landmarks~\cite{HolmesArditi1998, RowellUngar2005}. Our work with video games necessitates a digital solution; as such, the smartphone map interface we implemented also derives from prior work in touchscreen-based accessible graphics, particularly in presenting floor plans and other maps to VIPs~\cite{Su2010, Gotzelmann2015, Goncu2011, Goncu2015, Giudice2020Maps}.

When a player places their finger on the screen, they will begin surveying at their position, regardless of where on the screen they are touching. As they move their finger, they will survey the map relative to their  position, with the app announcing anything that the player touches. The app will announce all items in the world (as well as the player's position) using sound effects and/or text-to-speech. The app only reacts to touches within the current room that the player is in --- if the player drags their finger outside the room, a continuous warning tone will play.

In the first version of this tool, players started surveying at the portion of the map where their finger touched the screen; however, our visually impaired pilot participants ended up spending large amounts of time searching for their current position, which frustrated them. As a result, our second and final version registers a player’s initial touch at their current position.

\subsection{Whole-Room Shockwave}
\label{sec:toolshockwave}

The whole-room shockwave, depicted in the upper-right corner of Figure \ref{fig:tools}, uses an acoustic shockwave that the player triggers to communicate information about their surroundings. When the shockwave hits anything in the room, an announcement and/or sound effect emanates from that object via 3D sound. The shockwave corresponds to real-world physics in that closer objects will emanate their sounds back to the player before objects that are further away. If the player moves while the shockwave is active, the rate of expansion will match the player's speed.

The whole-room shockwave originated from our explorations in echolocation, which has been shown to promote spatial awareness in VIPs by communicating physical properties of the room and nearby objects~\cite{Kish2009, Norman2021, ThalerGoodale2016}. Our initial echolocation prototype had players press a button on their game controller to emit a click sound originating from the player’s position, similar to how some VIPs use echolocation within the physical world~\cite{Kish2009, Kolarik2014}. Our echolocation prototype was similar to virtual echolocation techniques used in prior work~\cite{Andrade2018} that used Steam Audio’s built-in head-related transfer function~\cite{ValveSoftware2019} to generate sound reflections based on the physical structure of each room.

In our pilot tests, however, we found that echolocation by itself was not equivalent to the other tools. While echolocation communicates the raw layout of an area, the other tools can communicate raw layout \textit{in addition to} specific object information through sound effects and text-to-speech. Furthermore, our visually impaired pilot participants were not at all experienced in echolocation and did not know how to decode and interpret the sound echoes in our game environment; they could only interpret broad qualities of the area such as how large it was. Although users could learn to use echolocation, prior work has indicated that it may take weeks for users to learn how to use click-based echolocation effectively~\cite{Norman2021}.

We, thus, made modifications to the initial \textit{echolocation} design and created the \textit{whole-room shockwave} --- a refined and more comprehensible version of echolocation. In its first iteration, the shockwave announced \textit{every} item that it hit, which proved to be auditorily overwhelming. Furthermore, both visually impaired participants found the shockwave to be too fast. As a result, our second and final version halved the speed of the shockwave and implemented a filtering mechanism. Players can press the right button on the D-pad to cycle through four filtering options --- all objects, mission-critical points-of-interest, non-mission-critical (decorative) objects, and walls only. Only items within the selected category will emit sounds during a shockwave.

\subsection{Directional Scanner}

The directional scanner, illustrated in the lower-left corner of Figure \ref{fig:tools}, allows players to survey in any direction using the right thumbstick. Players use the tool by tilting the thumbstick in any direction. This triggers an announcement naming the first object that lies in that direction via line-of-sight with respect to the player’s current position and orientation. The announcement is made via 3D sound from the point of the object in space. If the first object in a direction being pointed at is \textit{not} an object of interest (i.e., a wall or other generic obstruction), the scanner will emit a 440 Hz sine tone from the direction of the obstruction.

This tool represents prior work that has sought to replicate the act of ``looking around'' (or \textit{directionally scanning} an area) to promote spatial awareness for VIPs. We take particular inspiration from NavStick~\cite{Nair2021pp, NairSmith2020}, which introduced the concept of directional scanning within game worlds and showed how VIPs enjoyed the ability to survey their game environments directly by ``looking around.’’ Some prior work with directional scanning also exists in the physical world. Talking Points 3~\cite{Yang2011} is one such example: It features a ``Directional Finder'' that provides a list of landmarks that lie in the general direction that a VIP points their mobile device.

Our implementation of the directional scanner was derived from NavStick, and we did not implement any major changes to it as a result of our pilot tests.

\subsection{Simple Audio Menu}

The simple audio menu, shown in the lower-right corner of Figure \ref{fig:tools}, represents the idea of using a list to promote spatial awareness --- in particular, by allowing VIPs to learn about the contents of the area they are currently in. Many audio games made for VIPs, such as \textit{Terraformers}~\cite{Westin2004, PinInteractive2003a} and \textit{A Hero's Call}~\cite{OutOfSightGames2017}, use list- and grid-based representations to present the world to VIPs.

The simple audio menu we implemented exposes an audio-based list of points-of-interest (POIs). Players use the tool by pressing the left bumper button to open a list of POIs within the room they are currently in. As the player scrolls through the list using the D-pad, they will hear each item’s associated sound effect and text-to-speech announcement. The simple audio menu is modeled after list interfaces used in some audio games as well as prior research~\cite{Westin2004, PinInteractive2003a, Nair2021pp} in that it employs an alphabetical ordering of items. Previous research has suggested that, for a linear menu, a stable alphabetical ordering is less confusing than a proximity-based or direction-based ordering, both of which can change as the player moves~\cite{Nair2021pp}.

Similar to the directional scanner, we did not implement any major changes to the simple audio menu as a result of our pilots.

%% file: sec04-userstudy.tex
We performed a user study to investigate two important research questions about SATs within video games for VIPs:

\begin{enumerate}[label=\textbf{RQ\arabic*:}, leftmargin=3.5\parindent]
    \item What aspects of spatial awareness do VIPs find important within games?
    \item How well do today's differing SAT approaches --- as represented by the four tools we implemented --- facilitate each aspect of spatial awareness, and why?
\end{enumerate}

We created a 3D adventure game called \textit{Dungeon Escape} to investigate these two research questions. We included the four tools within \textit{Dungeon Escape} and used the game to run a user study with VIPs. In this section, we describe \textit{Dungeon Escape} and our user study.

\subsection{Game: \textit{Dungeon Escape}}

\textit{Dungeon Escape} is a 3D third-person adventure game set in a fantasy world, in which the player must escape small dungeons by finding objects that allow them to clear obstacles. We chose to create \textit{Dungeon Escape} to address RQ1 and RQ2 because the game requires players to use the tools they are given to search for and understand where objects are located and how the rooms in each level are laid out in order to succeed --- thus testing how well they are able to gain spatial awareness using those tools.

\begin{figure}[]
  \centering
  \includegraphics[width=0.47\textwidth]{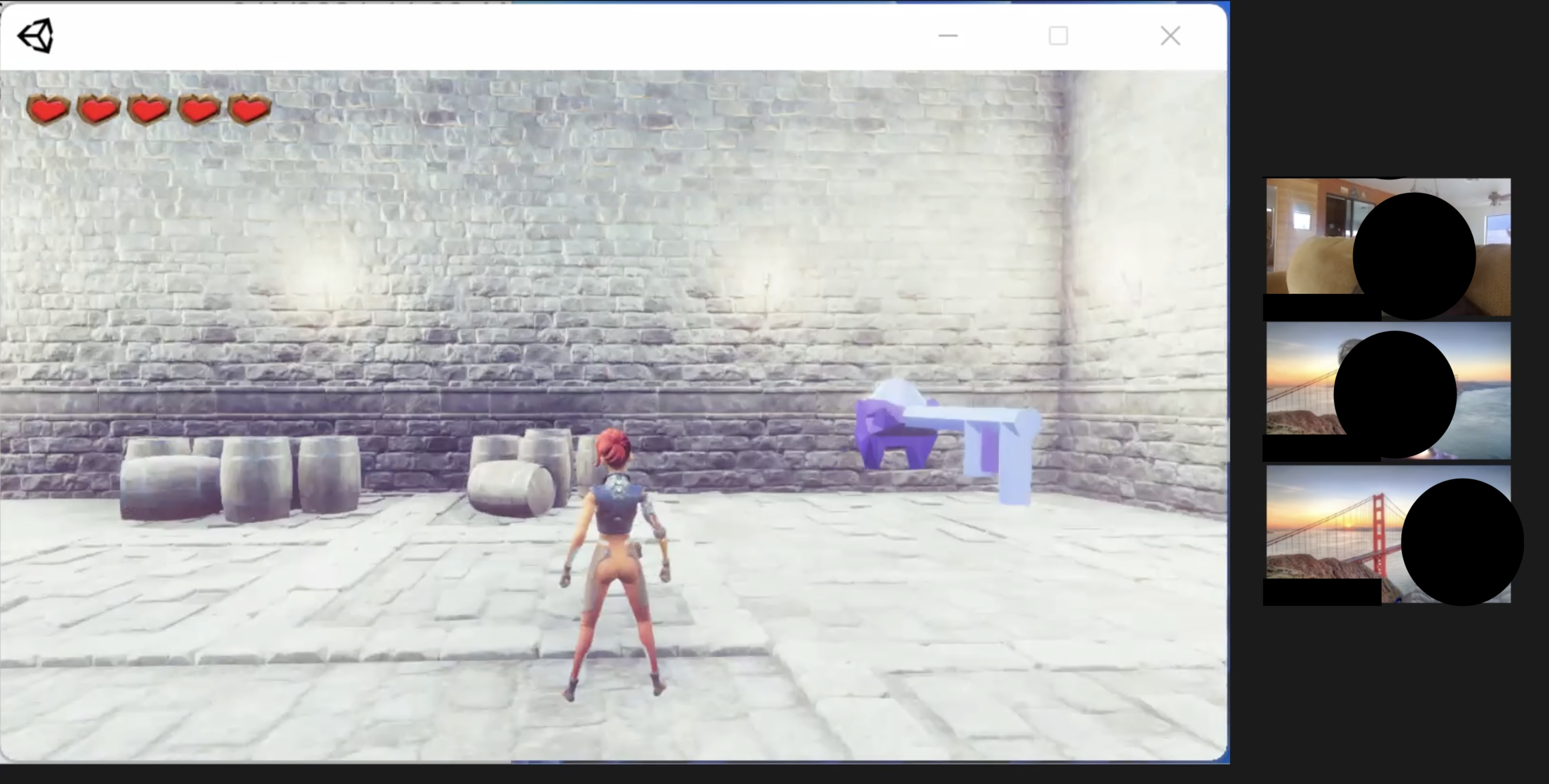}
  \caption{Remote study session with a participant and two facilitators. The participant is currently sharing their screen. Within the game, a blue key is situated to the participant's right. The participant will need to collect that key to progress through the level. \textit{(Faces obscured to protect anonymity.)}}
  \Description{Screen capture from a remote study session. A view of Dungeon Escape as a participant plays through it is alongside three obscured faces.}
  \label{fig:zoomsession}
\end{figure}

We created \textit{Dungeon Escape} using the Unity game engine~\cite{UnityTechnologies2020}, and we designed the dungeon’s layout using the Dungeon Architect Unity asset~\cite{DungeonArchitect}. Figures ~\ref{fig:tools} and ~\ref{fig:zoomsession} show views from \textit{Dungeon Escape}.

\begin{figure}[]
  \centering
  \includegraphics[width=0.47\textwidth]{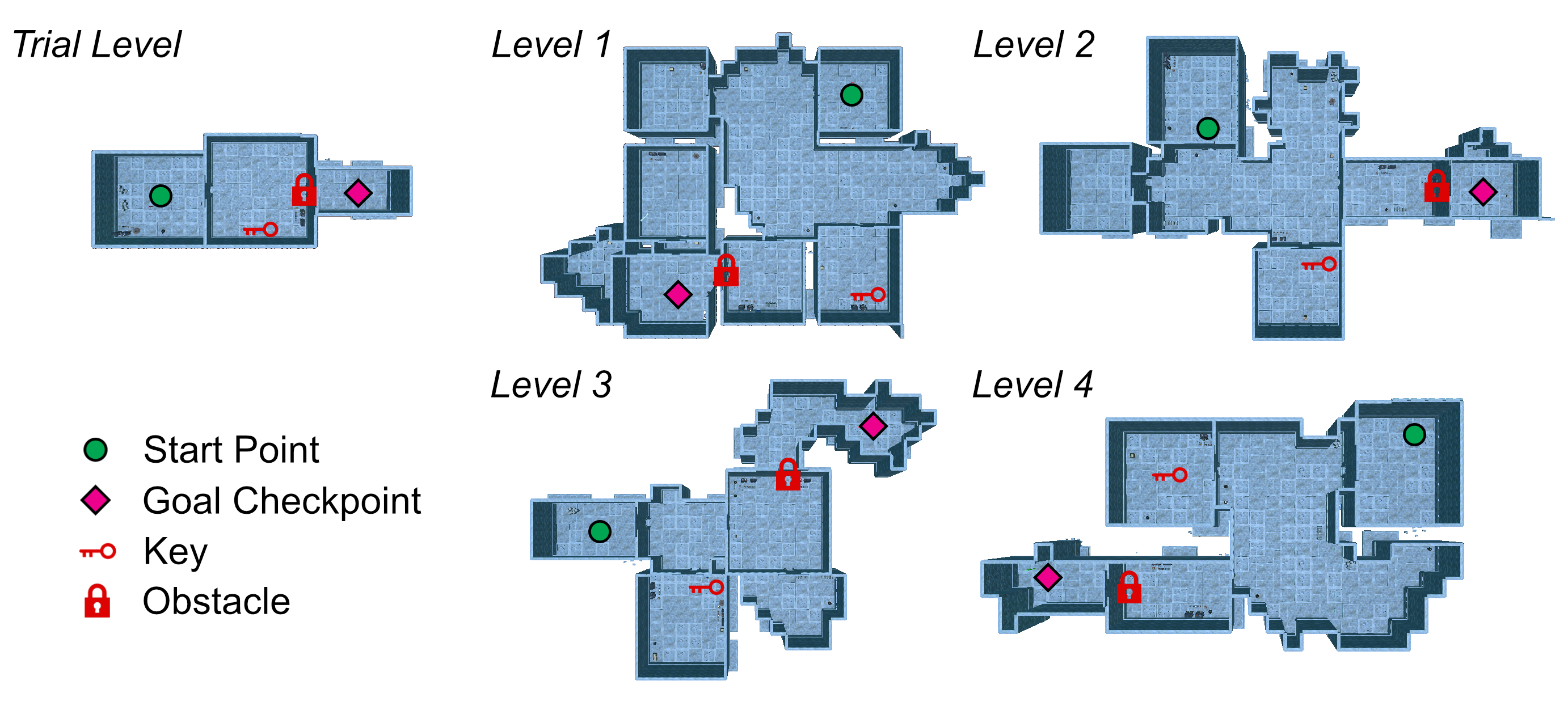}
  \caption{Overhead views of \textit{Dungeon Escape}'s trial level and four main levels. All levels have a common set of points-of-interest: a start point, a key, an obstacle that the key affords passage through, and a goal checkpoint. Each level, however, possesses a unique layout, allowing us to evaluate the four spatial awareness tools within a variety of layouts.}
  \Description{Bird’s eye views of five different room layouts. The four main levels each consist of 5 to 7 rooms with two larger “halls.” The trial level only consists of 3 rooms.}
  \label{fig:leveloverhead}
\end{figure}

\textit{Dungeon Escape} consists of four levels (small dungeons), which allowed us to study the four SATs within separate dungeon layouts. Figure \ref{fig:leveloverhead} shows overhead views of all four main levels and the trial level. In each main level, the player must reach a goal area by gaining passage through an obstacle: either a locked door, a cracked wooden door, a spider web, or a dog blocking the exit. To do so, the player must find a relevant object in another room: a key, an axe, a burning torch, or a bone, respectively. Each level consists of several rooms scattered with decorative objects such as crates and barrels.

We generated the four level layouts by deriving them from a single Dungeon Architect ``grid flow.'' This grid flow defined basic parameters from which Dungeon Architect would generate levels. Each level consisted of:

\begin{itemize}
    \item A ``start room'' within which the player first spawns.
    \item An ``obstacle room'' containing the obstacle.
    \item A ``key room'' containing the object that clears the obstacle.
    \item A ``main hall'' connecting the start, key, and obstacle rooms.
    \item A ``final hall'' containing the goal checkpoint.
\end{itemize}

We then fed random seed values into this grid flow to generate the final layouts. This allowed us to have unique level layouts while keeping them equivalent in terms of difficulty and structure. The trial level followed a similar conceptual structure but was much smaller, consisting of a start room, a combined key-and-obstacle room, and a final hall with the checkpoint.

Players move the main character with the left thumbstick. Tilting it forward and backward will move the character forward and backward. Tilting it left and right will rotate the character left and right. This control scheme reflects controls found in mainstream 3D games such as \textit{Tomb Raider}~\cite{TombRaider}, \textit{Resident Evil} 1-5~\cite{ResidentEvil1, ResidentEvil2, ResidentEvil3, ResidentEvil4, ResidentEvil5}, \textit{Metroid Prime} 1-3~\cite{MetroidPrime1, MetroidPrime2, MetroidPrime3}, \textit{Heavy Rain}~\cite{HeavyRain}, and \textit{Silent Hill}~\cite{SilentHill}, which use a fixed over-the-shoulder camera and use left/right on the left thumbstick to rotate the character. The right thumbstick is used by the directional scanner condition; thus, to eliminate a confound, we removed right thumbstick controls from all other conditions. Players can press the bottom face button to pick up an object or to use an object to remove the relevant obstacle.

Players hear a scraping sound if they physically hit an obstruction; the sound will be situated in the direction of contact. Keys, obstacles, and checkpoints play a relevant sound once the player is within two meters of the object. Players will hear the name of the room (for example, ``Start Room'' or ``Key Room'') announced on entry, and they can also press the right face button to hear the room name on-demand. We integrated these sounds to allow VIPs to be informed of these events --- i.e., hitting a wall or entering a room --- when they occur. Sighted players can perceive these events solely via sight, but VIPs require notifications via other means.

We also implemented a ``rotation indicator'' utility that helps players understand how much they are rotating when they turn left or right using the left thumbstick. As the player rotates, a click sound will be played at $15\degree$ increments via 3D sound \textit{only} in the direction of the player’s objective (i.e., the obstacle that must be cleared). The rotation indicator mimics snap rotation controls found in many games created for VIPs~\cite{Westin2004, PinInteractive2003a, Kaldobsky2011}, which allow players to snap to pre-defined angle increments. In order to bring \textit{Dungeon Escape}'s controls closer to the free movement of mainstream 3D games, we gave players full analog control via the left thumbstick but maintained the feedback afforded by snap rotation via the rotation indicator. The rotation indicator was available across all four tools and pointed in the direction of the objective regardless of any intervening obstacles. Similar utilities have also been implemented in prior work that has investigated navigation by VIPs within virtual environments~\cite{Andrade2018, Andrade2021, Nair2021pp}.

Additionally, players could place looping audio beacons on objects of interest so that they could lock onto and keep targets within their “field of view.” Once placed, these beacons emit a looping sound, which players can use to orient themselves and move towards the target. With NavStick, players point at a target with the right stick and press the right bumper button to place the beacon. With the simple audio menu, players scroll to a target and press the left bumper. With the smartphone map, players tap on the upper one-fifth of the screen to place a beacon at the last announced target. There was no mechanism for beacon placement with the whole-room shockwave. We added the beacons exclusively for guidance purposes to speed up the process of walking toward a target --- players still need to use an SAT to find objects and other targets before they can place a beacon at that object/target.

\subsection{Participants}

We recruited nine participants for this study. In our pre-study questionnaire, eight described themselves as being completely blind and one (P1) described themselves as having light perception only. All participants were male and have had their vision impairments from birth. Six participants were 18--25 years old; two (P5 \& P9) were 26--35 years old; and one (P3) was 36--45 years old. In addition to having vision impairments, two participants (P3 \& P6) reported having slight hearing loss in one of their ears.

We recruited participants \replaced{through posts on the AudioGames.net Forum, an online discussion board that centers around audio-based games and is frequented by VIPs.}{by posting to online forums popular among the visually impaired community \textit{(exact forums anonymized for submission)}.} Six of our participants reported themselves as being very experienced with video and other electronic games (4+ on a 5-point Likert scale), while the other three (P2, P6, \& P8) reported themselves as being moderately experienced with games (3 on a 5-point Likert scale).

\subsection{Technical Setup \& COVID-19 Challenges}

We performed this study remotely due to the COVID-19 pandemic and the difficulties that VIPs may face in travelling to our institution. We sent each participant an executable of our game for them to download to their computer before their study appointment. The game included all of the SATs except for the smartphone map. We distributed that tool as both iOS and Android apps using the Google Firebase App Distribution service~\cite{FirebaseAD}. We designed both \textit{Dungeon Escape} and the smartphone map to connect with a cloud backend, which allowed both components to synchronize with each other, and allowed us to remotely observe and control the runtime state of participants’ games using a custom-built control panel. 

We held the study appointments over Zoom and asked participants to share their computer audio (and, optionally, video) with us. Although there was no way for us to see the smartphone map during the study, most participants’ microphones picked up the sound from the app. Figure \ref{fig:zoomsession} shows a study session in progress. The study and our data collection efforts were approved by the Columbia University Institutional Review Board (IRB).

\subsection{Procedure}
\label{sec:procedure}

To address RQ1, we began the session by administering a two-part pre-study questionnaire. The first part requested demographic information alongside information about participants' existing experience with video games and physical world navigation. The second part directly asked participants about how important they find each of the six types of spatial awareness --- that we identified in Section \ref{sec:affdesc} --- within a video game context. For each type, responses were given on a 5-point unipolar Likert scale where 1 indicated that the type of spatial awareness was not-at-all important and 5 indicated that it was extremely important. Afterwards, we placed participants in a room within the game where we introduced basic movement and interaction controls.

For each tool, we first placed participants into the trial level. We explained how to use the tool and afterwards allowed participants to traverse the trial level at their own leisure. The trial level was the same across all tools. After the trial level, we placed participants into one of the four main levels. Although all participants played the four levels in the same order, we counterbalanced the order of the tools themselves via a Latin square design to reduce any variations caused by order effects.

In order to address RQ2, we administered a two-part post-level questionnaire; we did this after participants traversed a level with a tool. In the first part, we asked participants to elaborate on their impressions of the tool, what they think is missing, and in what game situations they might use the tool. In the second part, we gauged how well participants thought the tool satisfied each of the six types of spatial awareness. Responses were given on five-point scales, where 1 indicated that the tool facilitated that type not-at-all well and 5 indicated that it facilitated that type extremely well. Participants were encouraged to elaborate on all questions.

After completing all four levels, we administered a two-part post-study questionnaire. In the first part, we asked participants to consider a scenario where they were able to play the levels in \textit{Dungeon Escape} using more than one tool at once; we did this in order to determine if using multiple tools at once could have improved participants' spatial awareness in any way. As part of this section, participants were asked to provide two combinations of two tools each that they would have liked to use if they were given the chance to do so. (We should note that \textit{Dungeon Escape} is capable of activating two tools at once; however, in our initial pilot tests, including additional game levels to test these combinations made study sessions well exceed our limit of two hours.) In the second part, we asked participants how likely they were to recommend each individual tool to a friend or colleague, assuming they had the same visual impairments as the participant. Responses were given on a 10-point net promoter score scale~\cite{Reichheld2003}, where 1 indicated they were very unlikely to recommend it and 10 was very likely. 

\subsection{Data Collection \& Analysis}

We administered all questionnaires by having the facilitator read out each question and input the participant's response into an internal Google Form. For all choice- and rating-based questions, we asked for participants' open-ended opinions via the questionnaire itself by explicitly following up on their responses. The facilitator was also encouraged to follow up on any other points they found interesting throughout the session --- though they were not allowed to disturb the participant while a game level was in progress. We have included the questionnaires as part of our supplementary material.

We recorded all sessions with participants' permission for transcription purposes. We also obtained raw data of participants' actions within the game by capturing in-game logs.

To analyze sessions, we followed an inductive coding process that involved five members of the research team. Individual coders went through session transcripts and coded quotes and other events. Then, all five coders iterated on the codes together until there was unanimous agreement that they could not iterate further.

%% file: sec05-RQ1results.tex
\begin{figure}[]
  \centering
  \includegraphics[width=0.47\textwidth]{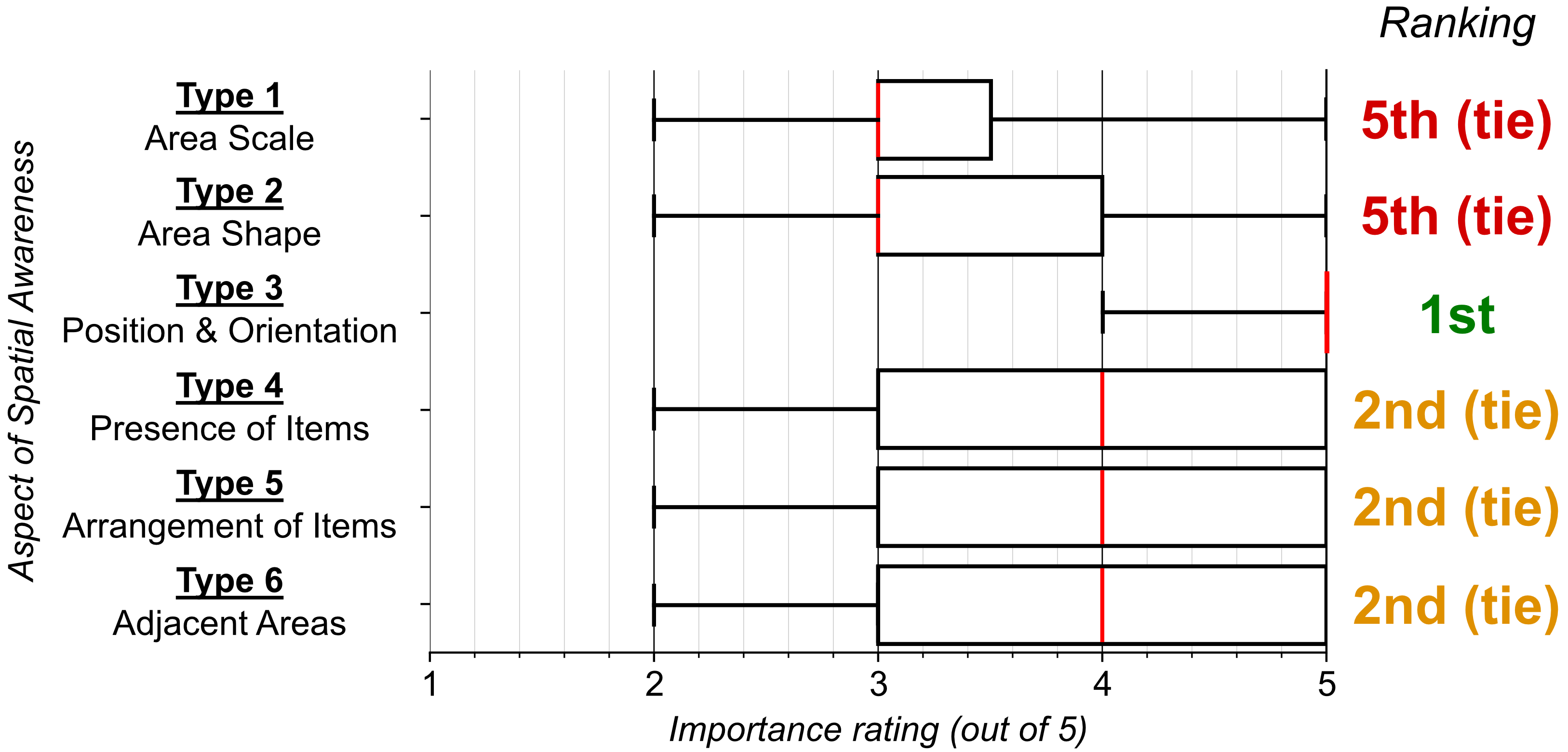}
  \caption{The importance of the six aspects of spatial awareness within games for VIPs (RQ1). Responses were given on a five-point unipolar Likert scale. Red lines indicate median ratings within this box plot. Rankings based on median ratings are shown to the right.}
  \Description{Box plot showing importance ratings for each spatial awareness type by VIPs. Type 3 had a median importance rating of 5 out of 5. Types 4, 5, and 6 had medians of 4. Types 1 and 2 each had medians of 3.}
  \label{fig:affvalue_box}
\end{figure}


In this section, we report our findings regarding our first research question (RQ1): learning what aspects of spatial awareness VIPs find important \textit{within games}. We captured these opinions in the second questionnaire that we administered as part of our pre-study procedure. 

Figure \ref{fig:affvalue_box} shows box plots of participants' importance ratings for the six types of spatial awareness. Note that the median rating for each measure is highlighted in red in each type's box plot.

The data suggest that the six types of spatial awareness can be divided into three levels of importance for VIPs. The \textit{most} important aspect of spatial awareness to VIPs within games is position and orientation (Type 3) awareness, which received a median importance rating of 5 on a 5-point unipolar Likert scale. Below that, three spatial awareness aspects --- presence of items (Type 4), arrangement of items (Type 5), and adjacent areas (Type 6) --- tied each other with a median importance rating of 4. The \textit{least} important aspects of spatial awareness to VIPs within games are scale (Type 1) and shape (Type 2) awareness, both of which received median importance ratings of 3.

The following subsections dive deeper into participants' reasoning behind how important they rated each aspect of spatial awareness to be within games. All quotes come from participants' open-ended responses while completing this questionnaire. In Section~\ref{sec:discussion_section}, we discuss how these findings and the findings from RQ2 (in Section~\ref{sec:toolresults}) collectively reveal new design considerations and research opportunities for spatial awareness tools.

\subsection{Rank 1: Position and Orientation [Type 3]}

Participants found position and orientation to be the most important aspect of spatial awareness within a video game context. Six participants explicitly affirmed this aspect of spatial awareness as the most important because it was crucial to determining their current state within the game world:

\begin{quote}
    \textit{“You have an idea of how fast you’re turning and in what direction. I would say it’s the most important thing.”} --- \textbf{P3}
\end{quote}

Another participant who echoed this sentiment, P9, recounted extensive experience with shooter audio games, such as \textit{Swamp}~\cite{Kaldobsky2011}, that require players to move through a complex environment. P9 affirmed position \& orientation awareness --- and thus, awareness of their current state --- as extremely important to helping them plan out future actions, which is a crucial aspect of shooter-type games:

\begin{quote}
    \textit{“You have to know where you are at and where you are oriented to in order to know where to go and what to do next.”} --- \textbf{P9}
\end{quote}

These opinions reflect work within the physical world that has found position and orientation to be important to VIPs~\cite{GiudiceLegge2008, Giudice2020}. They also establish that SATs for VIPs within video games must satisfy a high bar in terms of communicating position and orientation information. In Section \ref{sec:toolresults}, we determine if any of the four tools we implement for this study satisfy this high bar.


\subsection{Rank 2 (three-way tie): Presence, Arrangement, and Adjacent Areas [Types 4, 5, and 6]}

After position and orientation, participants found the next most important aspects of spatial awareness to be the presence and arrangement of items within the space (Types 4 \& 5) and information about areas adjacent to their current area (Type 6). 

These aspects are all important to participants in certain contexts, but not in all situations like position and orientation is. As P3 and P9 implied in their quotes in the previous subsection, position and orientation awareness grants players with a sense of their \textit{current state} within the world. Prior work in the physical world has found ascertaining this knowledge to be cognitively demanding for VIPs as they move through an environment~\cite{GiudiceLegge2008, Lewis2015}. This increased cognitive load can interfere with VIPs' ability to understand \textit{other} aspects of spatial awareness, making position and orientation awareness essential.


Seven participants found presence to be very important for spatial awareness because if an SAT did well at communicating presence, then they could be confident that they would not miss finding anything within the game:

\begin{quote}
    \textit{“If you hear [a familiar object], you know you are relatively in the right place and you can search the area specifically.”} --- \textbf{P7}
\end{quote}

Five participants clarified why the importance of knowing the arrangement of items is heavily context-dependent. For example, when faced with objectives that involve finding a specific item, participants believed that knowing the arrangement of items was very helpful because it would help them figure out where to go first. However, participants noted that having knowledge about items' arrangement may be detrimental in less restrictive, exploration-oriented tasks since that knowledge may reveal too much information and rob players the enjoyment of discovering items for themselves:

\begin{quote}
    \textit{“[Knowing arrangement] depends on what the task is [at hand]. It's especially [important] if it involves triggering certain things in certain orders, finding an item then finding a person, or facing off against a challenger then finding an NPC.”} --- \textbf{P7}
\end{quote}

Six participants felt that having an SAT communicate which regions are adjacent to the one they are currently in would be beneficial as it would make navigation through the game world easier:

\begin{quote}
    \textit{“To be honest, I’ll say [having an SAT communicate adjacencies is] extremely important because it makes it much easier for the player to move from one area to another without moving through the whole map.”} --- \textbf{P5}
\end{quote}

However, three others feared that having this information presented outright may make exploration and discovery less fun. One such participant was P4 who was a fan of games that required a high level of strategy. P4 asserted that the game should preserve a level of challenge and instead convey connections to other navigable places using plot and contextual cues such as dialogue or readable signs in the world itself:

\begin{quote}
    \textit{“If it's one of those strategy games where you have to discover it on your own, it’s not important. Let’s say it’s a hidden area, [...] it should stay hidden.”} --- \textbf{P4}
\end{quote}

The other two participants, P2 and P6, echoed similar sentiments. 

This finding is quite surprising: These three participants thought that an SAT did not necessarily need to communicate adjacencies \textit{despite} us identifying this as a basic aspect of spatial awareness. This points to the importance that participants place in their \textit{experience} within the game over the actual \textit{information} they receive, implying that VIPs may be willing to sacrifice receiving some pieces of information for the sake of a more interesting gaming experience.

\subsection{Rank 5 (two-way tie): Shape and Scale [Types 1 \& 2]}

Participants generally found scale and shape information about an area to be the least important aspects of spatial awareness. VIPs' opinions generally revolved around the sentiment that, unlike the other types of spatial awareness, scale and shape information may be outright unnecessary much of the time.

Seven participants stated that having a sense of the room's scale was not important to them and that SATs should focus on conveying information about the presence and location of nearby objects instead:

\begin{quote}
    \textit{“I feel when you are navigating in games you don't really need to know how big the area is as long as you know where the objects in that area are.”} --- \textbf{P1}
\end{quote}

Seven participants thought that an SAT should only convey information about the surrounding area’s shape when absolutely necessary and that communicating shape information is too much for an SAT to do, possibly resulting in information overload. However, these participants also thought that knowing shape information in some situations may make navigation more efficient --- for example, in a situation where the room does not have a circular or rectangular shape:

\begin{quote}
    \textit{“If the room is an odd shape --- every time I play a game, I assume the room is like a square, but that’s not always the case, sometimes rooms may have [many] sides, parts that jut out --- so I believe it’s a consideration.”} --- \textbf{P3}
\end{quote}

In a situation where a room is not rectangular, knowing the room’s shape could help players plan out their movements more carefully. Otherwise, players may resort to hugging walls to traverse the room and search for doors, which may become frustrating.

%% file: sec06-RQ2results.tex
\begin{figure}[]
  \centering
  \includegraphics[width=0.47\textwidth]{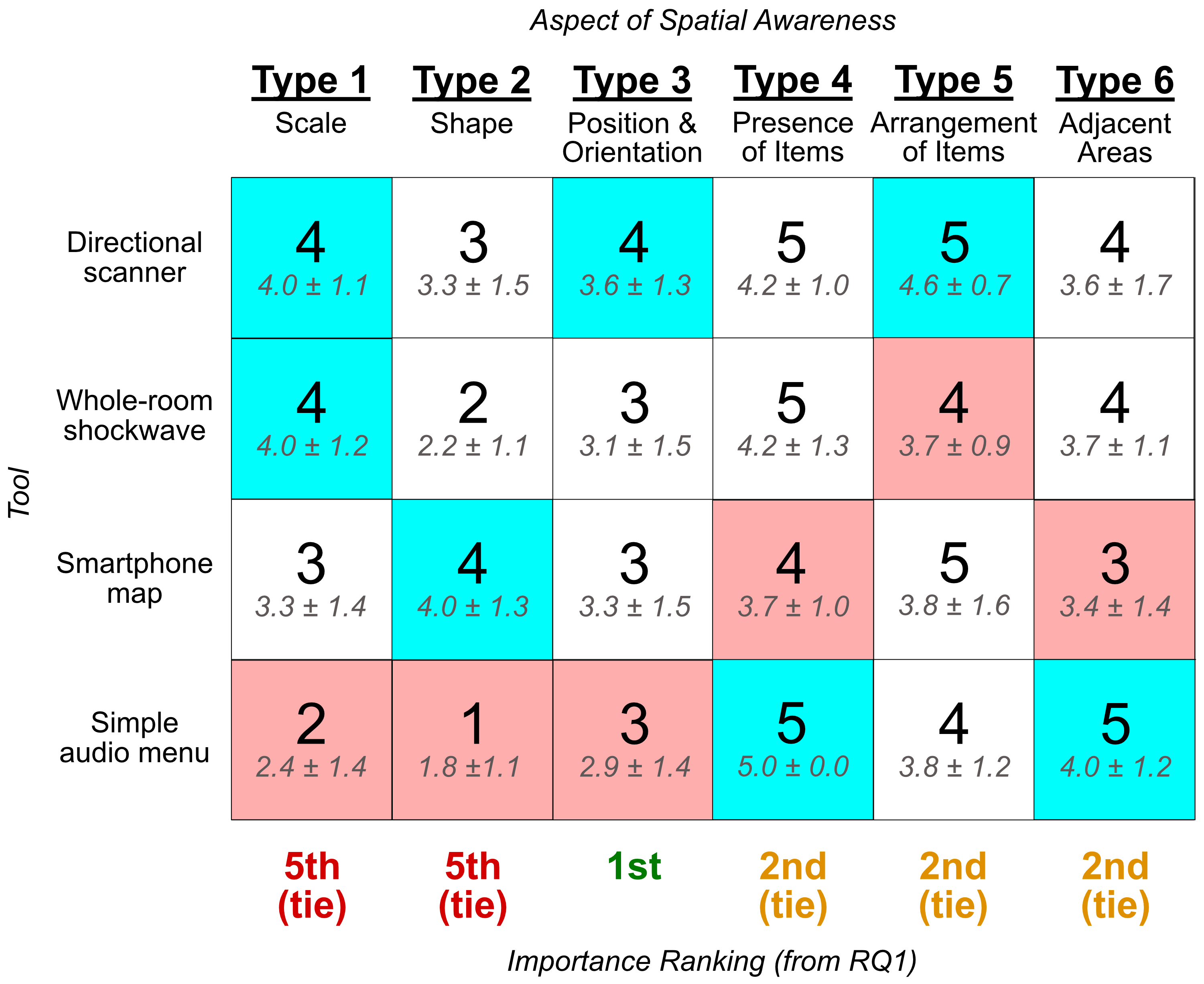}
  \caption{Participant ratings for how well each SAT approach facilitates each type of spatial awareness. Large text indicates median values, and small text indicates mean plus/minus standard deviation. Responses were given on a five-point unipolar Likert scale. Blue and red cells indicate the best and worst performing SAT approaches for each spatial awareness type (column), respectively. The importance rankings from our RQ1 analysis are shown again at the bottom.}
  \Description{Table showing median and mean participant ratings for each tool and each aspect of spatial awareness. For scale, the directional scanner and the shockwave had the highest score (4 out of 5); the simple audio menu had the lowest (2). For shape, the smartphone had the highest (4) and the simple menu had the lowest (1). For position & orientation, the scanner had the highest (4), and the simple menu had the lowest (1). For presence, the simple menu had the highest (5), and the smartphone had the lowest (4). For arrangement, the scanner had the highest (5) and the shockwave had the lowest (4). For adjacencies, the menu had the highest (5) and the map had the lowest (3).}
  \label{fig:heatmap}
\end{figure}


In this section, we report our results regarding our second research question (RQ2): determining how well the four tools we implemented facilitate the various aspects of spatial awareness. Combined with our RQ1 results, these results will shed light on how SATs should be designed to best facilitate the aspects of spatial awareness that are most important to VIPs.

Figure \ref{fig:heatmap} shows an overview of the ``winners'' and ``losers'' in terms of participants’ post-level responses on how well each tool facilitated the various aspects of spatial awareness. We determined these by looking at the median ratings for all four tools for a given spatial awareness aspect. A tool ``wins'' an aspect of spatial awareness if it has the highest median rating out of all the tools; a tools ``loses`` an aspect of spatial awareness if it has the lowest median rating. We broke any ties using the \textit{mean} rating.

We see that the directional scanner scored the highest in terms of facilitating three aspects of spatial awareness (scale, position \& orientation, and arrangement), meaning that participants thought it was the best tool for facilitating those aspects. The simple audio menu scored the highest in two types (presence and adjacencies), and the smartphone interface and whole-room shockwave scored the highest in one type each (area shape and tied with the directional scanner on area scale, respectively). The directional scanner did not score the \textit{lowest} on any type.

The results presented in this section are organized thematically, with each theme representing opinions shared by a majority of our participants.

\subsection{Since participants could trace out the contours of a room with their finger, the smartphone map communicated the shape of an area better than other tools.}
\label{sec:spmapshape}

Figure \ref{fig:heatmap} shows that participants’ ratings on how the smartphone map communicated the shape (Type~2) of the room were generally the highest out of all four tools. Five participants resonated with the following sentiment:

\begin{quote}
    \textit{"[The smartphone map] gives you a general idea of where openings and spaces are. It [also] gives you an idea of how far each edge is from your center point which tells you, ‘OK, [the wall] angles a bit.’"} --- \textbf{P6}
\end{quote}

One participant even used this ability to their advantage. For example, in the irregularly-shaped final room of Level 3, P8 used the smartphone interface to trace out the walls of the room and determine that they needed to turn a corner to reach the goal checkpoint:

\begin{quote}
    \textit{“Having that memory of ‘Oh, I know a little bit more about the shape of the room than I had previously with the other tools.’ --- that really helped me get a better sense of exactly where I needed to go in terms of [knowing that] I have to round a corner instead of trying to run directly forward for the target.”} --- \textbf{P8}
\end{quote}

\subsection{The whole-room shockwave allowed participants to quickly ascertain a general overview of an area, especially with respect to its scale.}

Figure~\ref{fig:heatmap} shows that the whole-room shockwave tied the directional scanner for being the best tool at communicating an area's scale (Type~1). Participants found that the distance-based volume attenuation afforded by \textit{Dungeon Escape}'s 3D sound system and the delayed timing of objects’ sounds during a shockwave helped them approximate how far away objects were and, thus, how big the room was. 

P8 was one such participant; they described themselves as "not-at-all experienced" with echolocation techniques in the pre-study questionnaire. (Recall from Section \ref{sec:toolshockwave} that we derived the whole-room shockwave from our explorations in echolocation.) Yet, they relayed the following positive sentiment which was shared by many other participants despite their inexperience with echolocation: 

\begin{quote}
    \textit{“Even though doors and objects were further away from me, I was still able to know that they are still in fact there. [The shockwave] helped me quickly gauge 'OK, cool. I know I'm in a corridor [...] and I know there is a door in the far end, and so this helps me determine on a higher level how big the room might be.'”} --- \textbf{P8}
\end{quote}

The quick nature of the shockwave particularly advantaged participants within areas with many items. One such area within \textit{Dungeon Escape} was the irregularly shaped room mentioned in Section \ref{sec:spmapshape}, which contained obstacles in the form of barrels and crates. Participants who used the whole-room shockwave hit the checkpoint in that room much faster ($M =$ 19 sec., $SD =$ 2.8 sec.) than those who used the directional scanner ($M =$ 51 sec., $SD =$ 17.7 sec.), simple audio menu ($M =$ 123 sec., $SD =$ 5.0 sec.), and smartphone map ($M =$ 135 sec., $SD =$ 15.5 sec.). The shockwave provided participants with an almost-instant overview of what obstacles were in the room. However, those who used the other tools spent a much longer time searching for these very obstacles --- scrolling through each item (using the menu), trying to point at them (using the directional scanner), or trying to find them on a map (using the smartphone).

\subsection{Participants made extensive use of the whole-room shockwave's filters.}

All nine participants made use of the whole-room shockwave's filtering mechanism. Six of them explicitly mentioned that this ability was extremely important to them and that they highly valued this ability even in applications outside of games. One participant invoked the customizability of screen readers as an example:

\begin{quote}
    \textit{“Everybody has different needs and wants so I really believe in allowing information to be filtered in such a way where you just get the information you need as you need it like in the shockwave. [...] Screen readers have settings like this for a reason.”} — \textbf{P3}
\end{quote}

\subsection{The physical use of the right stick in the directional scanner meant that participants could obtain a clear idea of how items were arranged around themselves.}

Our findings with the directional scanner provided insights into how physically moving a joystick to survey an environment might provide players with an enhanced sense of its layout. Five participants explicitly mentioned that moving the joystick to ``look around'' allowed them to understand how objects were arranged:

\begin{quote}
    \textit{“Because of where I had to put the stick to see stuff around me, it really helped. It was easier to tell what was behind me, what was in front of me, or what was in any other direction because I knew where my stick position was.”} --- \textbf{P6}
\end{quote}

P6 went on to say that surveying with the joystick “felt natural” and compared the directional scanner to a camera which they could use to ``look’’ for objects. This sentiment is further reflected in Figure~\ref{fig:heatmap}, which shows that the directional scanner scored the highest out of all four tools in terms of communicating the arrangement of items (Type~5).

\begin{figure*}
    \centering
    \includegraphics[width=0.83\textwidth]{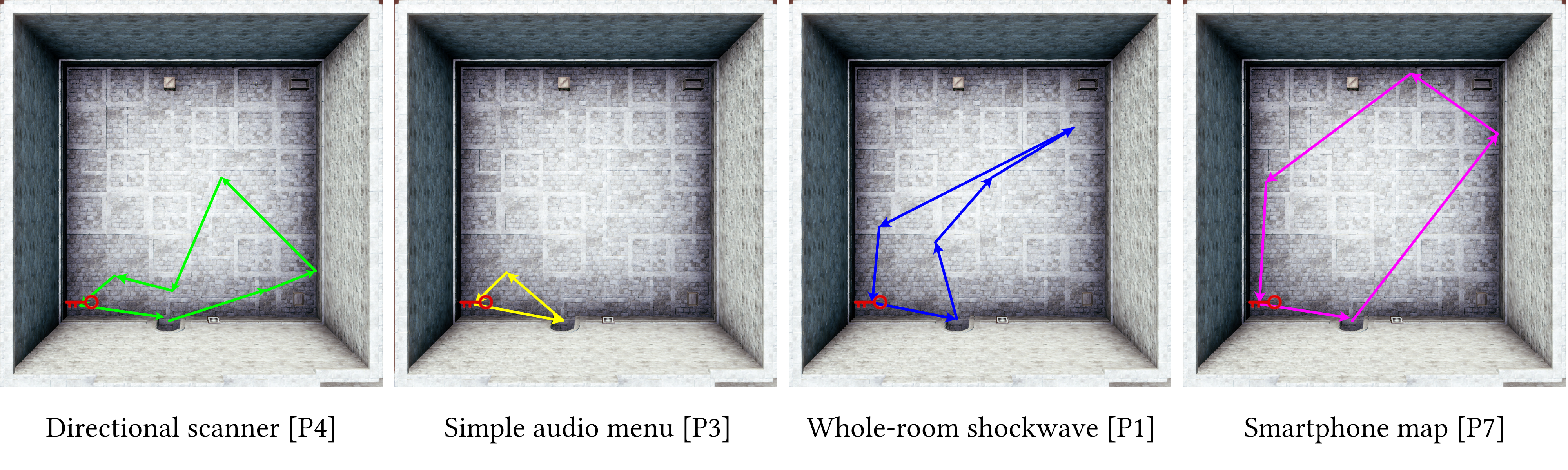}
    \vspace{-4mm}
    \caption{Illustration of paths taken by participants within the Level 2 key room. The subfigures depict four different participants traversing the same room with different tools; the key is located in the lower left corner, and participants enter through the door at the bottom. Paths are divided into segments --- the end of a segment represents a point where the participant paused to survey using the tool.}
    \Description{Four overhead views of the same room. Each view has a series of connected lines that represent a participant's path through the room while using the corresponding tool. The paths for the directional scanner, shockwave, and smartphone map cover greater areas of the room than that of the simple audio menu.}
\label{fig:pathplots}
\end{figure*}

\subsection{The simple audio menu’s straightforward presentation of items meant that participants received information about the presence of items extremely well.}

Participants found that the simple audio menu clearly communicated the presence of every item in the room. As seen in Figure \ref{fig:heatmap}, the simple audio menu received an average/median rating of 5.0 with a standard deviation of zero on communicating the presence of items (Type~4) --- every participant gave the menu the maximum possible score on this aspect of spatial awareness. Participants unanimously agreed that they were able to obtain a clear idea of what was in the room because of its straightforward presentation:

\begin{quote}
    \textit{“You know everything that’s there because it’s in a menu. There’s nothing hidden. It doesn't matter if you’re far away from it. If you’re in the same room as it, it's on that list. That’s something I really like.”} --- \textbf{P3}
\end{quote}

Yet, some participants complained that the simple audio menu provided \textit{too much} information. Within games, a certain degree of surprise and exploration --- that is, the ability to “discover” aspects of the game world --- are core elements for making a game fun for players~\cite{LeBlanc2008}. Knowing the presence of objects so easily can remove this aspect of discovery from the game. Indeed, five participants felt that the simple audio menu did not promote exploration and made the game less enjoyable:

\begin{quote}
    \textit{“I thought that I was using a shortcut. [...] I like that it’s faster, but it takes something out of the game experience.”} --- \textbf{P2}
\end{quote}

Within \textit{Dungeon Escape} itself, participants tended to avoid deviating from the task at hand while using the simple audio menu, going directly to POIs they needed to go to. Figure \ref{fig:pathplots} plots paths taken by participants within the key room in Level 2. Note how P3 went straight to the key when using the simple audio menu in Level 2, while participants who used other tools in the same level roamed around the room in an effort to survey their surroundings more thoroughly. This behavior is also visible in the raw time data we collected within the room: Those who used the menu collected the key much faster ($M =$ 17 sec., $SD =$ 4.2 sec.) than those who used the shockwave ($M =$ 84 sec., $SD =$ 6.5 sec.), smartphone ($M =$ 93 sec., $SD =$ 17.5 sec.), and directional scanner ($M =$ 105 sec., $SD =$ 5 sec.). Although players completed the levels with the simple audio menu (and often did so quickly), it remains an open question whether players' increased focus on objectives and lack of exploration is a net positive for the game experience or not.

\subsection{No tool excelled at communicating position and orientation.}
\label{sec:posorienttools}

As we found in Section~\ref{sec:affresults}, participants rated position and orientation (Type~3) as the most important aspect of spatial awareness to them. However, post-level ratings indicate that participants perceived all four tools to be mediocre at facilitating position and orientation information. As Figure~\ref{fig:heatmap} shows, the average score that each SAT received in terms of affording position and orientation information was a low-to-mid three (``moderately well''). Our results indicate that these four tools may not meet the high bar that these tools need to meet for such an important aspect of spatial awareness.

\subsection{Participants disliked having to juggle multiple pieces of hardware when using the smartphone map.}

Five participants mentioned that they found the smartphone map cumbersome to use. Participants often found themselves needing to physically switch between their controller and their smartphone when they wanted to explore the map. Furthermore, at least six participants used noise-cancelling headphones during their sessions and had to adjust \textit{them} as well to hear the smartphone's audio output. These experiences annoyed some participants:

\begin{quote}
    \textit{"I have mixed emotions [about the smartphone map] because I have to do one thing on one device and then move with the other device [...] That made things a bit confusing and annoying."} - \textbf{P2}
\end{quote}

\subsection{The simple audio menu did not communicate scale and shape well.}

As Figure \ref{fig:heatmap} shows, participants thought that the simple audio menu communicated the scale (Type~1) and shape (Type~2) of areas quite poorly with average ratings of around 2 out of 5. The simple audio menu did not explicitly communicate boundaries or other characteristics of the room itself. As such, many participants could not definitively determine the structure (scale and shape) of the surrounding area using the menu:

\begin{quote}
    \textit{"I could probably use [the simple audio menu’s] 3D sounds to assume that, say, a bunch of items were against a wall if they’re coming from the same general side relative to me [...] but that’s an educated guess."} --- \textbf{P6}
\end{quote}

\subsection{Participants found the whole-room shockwave to be overwhelming, which negatively affected their spatial awareness.}

Five participants felt that despite communicating scale relatively well, the whole-room shockwave provided too much information, which negatively affected their sense of spatial awareness. We were surprised by this finding since we designed the shockwave to emit slowly for better intelligibility, and every participant used the filters to make the shockwave easier to understand.

Yet, despite these improvements, participants still felt that the shockwave was too information-dense, making it difficult for them to ascertain information about their environment. Our conversations with them yielded insights into how VIPs view similar echolocation-inspired tools within other games. P3, who described themselves as having played "everything" when asked about their gaming experience during the pre-study, was especially vocal:

\begin{quote}
    \textit{“Everybody thinks you can just send out a sonar ping and get information about an environment. [...] Echolocation is very overwhelming, especially in a game. Trying to hone in on an item that is far away and being masked by another item is ludicrous.”} --- \textbf{P3}
\end{quote}

\subsection{Participants preferred combinations of tools that excelled across multiple spatial awareness aspects.}

Figure \ref{fig:heatmap} gives us an interesting perspective on how combinations of tools can best facilitate multiple aspects of spatial awareness jointly. As stated in Section \ref{sec:procedure}, we asked participants to state their two most preferred combinations of tools. The (directional scanner + simple audio menu) combination was one of the most selected combinations, with five participants selecting it. This combination "wins" in four out of the six aspects of spatial awareness: position \& orientation, presence of items, arrangement of items, and communicating adjacent areas. Furthermore, this combination is "tied" with the whole-room shockwave at being the best at conveying area scale. Five participants also selected the (directional scanner + whole-room shockwave) combination, which wins at \textit{three} of the six aspects of spatial awareness. We discuss these selections further in Section \ref{sec:discussion_section}.

%% file: sec07-discussion.tex
In Sections \ref{sec:affresults} and \ref{sec:toolresults}, we reported our findings about what aspects of spatial awareness VIPs find important (RQ1) and how well current SAT approaches facilitate the various aspects of spatial awareness (RQ2). In this section, we now synthesize these findings together to form broader takeaways for how SATs should be designed. We hope that game designers can use these takeaways to decide which SAT is best for them to incorporate into their game to make it accessible, if they only have the resources to implement one or two of them.

\subsection{Position and orientation is the most important type of spatial awareness for VIPs, yet is \textit{not} well-served by current tools.}
\label{sec:posorientdisc}

Our results indicate that communicating position and orientation well is a crucial challenge that must be addressed when designing future SATs. As we reported in Section \ref{sec:posorienttools}, participants rated position and orientation as the most important aspect of spatial awareness to them. Yet, they also felt that all four tools were mediocre at facilitating position and orientation information. Surprisingly, this includes the smartphone map, which was the tool that most explicitly communicated position and orientation information, as we described in Section~\ref{sec:toolmap}. This indicates that communicating position and orientation information to VIPs is harder than researchers assume and that a major opportunity for future research is to develop better indicators for VIPs' position and orientation.

Previous research has shown that VIPs rely heavily on landmarks and other environmental features to determine their position and orientation, which, in turn, allows them to navigate through environments~\cite{YuGanz2012, Fallah2012}. These landmarks include walls and other boundaries dictating the area's scale and shape (Types 1 and 2) as well as the layout of items within the space (Type 5). From our findings, however, we see that VIPs do not find it suitable to merely infer their position and orientation from these other cues and would rather benefit from having it communicated more explicitly.

\vspace{2mm}
\begin{center}
\setlength{\fboxsep}{0.8em}
\fbox{\begin{minipage}{0.4\textwidth}
\textbf{Design Implication \#1}: VIPs will benefit greatly from a purpose-built tool for communicating position and orientation in real time.
\end{minipage}}
\end{center}
\vspace{1mm}


\subsection{The four most important aspects of spatial awareness are covered by two tools.}
\label{sec:twotoolsdisc}

If we consider the four most important aspects of spatial awareness from our RQ1 findings --- position and orientation (Type~3), item presence (Type~4), item arrangement (Type~5), and adjacent areas (Type~6) --- we can see that the combination of the directional scanner with the simple audio menu ``wins'' at communicating all four of these types. We do not consider area scale (Type~1) and area shape (Type~2) because participants found them to be the least important; however, we can also see that the directional scanner is tied for ``winning'' scale as well. From a theoretical standpoint, this implies that VIPs would most gravitate toward this combination, and indeed, we saw precisely this during our study.

The fact that participants are excited about the (directional scanner + simple audio menu) combination makes sense. It seems that, with this combination of tools, participants gravitated toward a combination that facilitates the greatest number of spatial awareness aspects well.

\vspace{2mm}
\begin{center}
\setlength{\fboxsep}{0.8em}
\fbox{\begin{minipage}{0.4\textwidth}
\textbf{Design Implication \#2}: Of today's SATs, the combination of the directional scanner and simple audio menu gives VIPs the greatest spatial awareness.
\end{minipage}}
\end{center}
\vspace{1mm}


\begin{figure}[]
  \centering
  \includegraphics[width=0.47\textwidth]{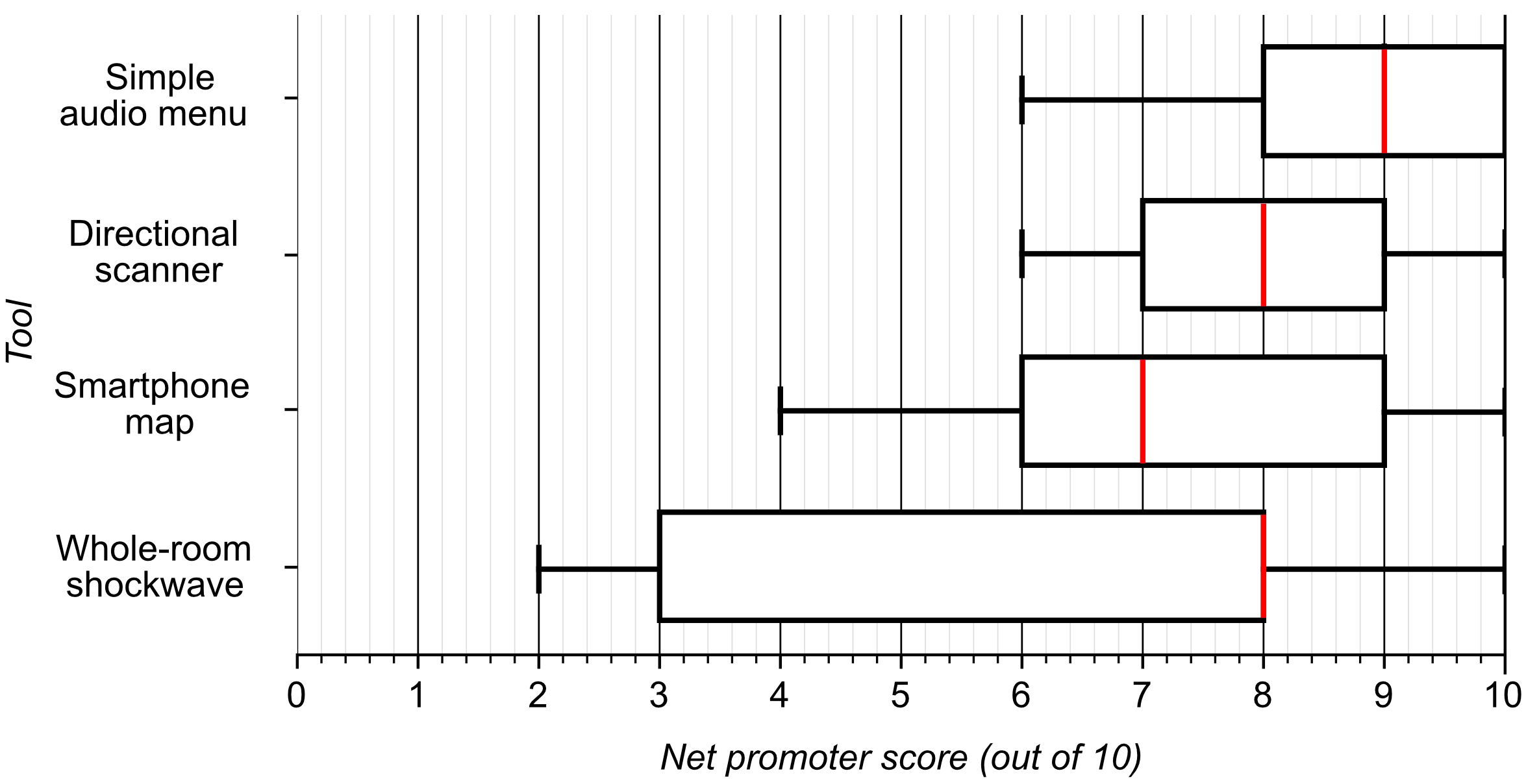}
  \caption{Box plot of net promoter score responses for all four tools. Red lines indicate median. The whole-room shockwave received some of the lowest scores out of all four tools.}
  \Description{Box plot of Net Promoter Score responses with respect to all four tools. The simple audio menu received a median score of 9; the directional scanner 8; the smartphone map 7; and the whole-room shockwave 8. The shockwave had the largest range of all four tools.}
  \label{fig:npsbox}
\end{figure}

\subsection{VIPs highly value the ability to customize SATs.}

In addition to the (directional scanner + simple audio menu) combination, five participants also picked the (directional scanner + whole-room shockwave) combination as one of their favorite combinations. Unlike the former combination, however, the latter combination only ``wins’’ at three of the six types of spatial awareness (i.e., the two tools cover winning values for three columns in Figure~\ref{fig:heatmap}). Additionally, Figure \ref{fig:npsbox} shows that the shockwave had some of the lowest net promoter scores out of all of the tools, and participants even complained about the tool being overwhelming. This finding implies that there exists a consideration that VIPs may find \textit{even more} important than raw spatial awareness. 

One possible explanation lies in the fact that the whole-room shockwave was the only tool that participants could change the behavior of --- in this case, selecting the type of information they wanted to hear. Participants' enthusiasm for customizable tools --- especially evident in their comparisons with other tools such as screen readers --- shows that SATs should implement similar capabilities, allowing VIPs to take control of what they hear. 

\vspace{2mm}
\begin{center}
\setlength{\fboxsep}{0.8em}
\fbox{\begin{minipage}{0.4\textwidth}
\textbf{Design Implication \#3}: SATs should embrace customizability, allowing VIPs to customize and filter the information communicated.
\end{minipage}}
\end{center}
\vspace{1mm}

%% file: sec08-futurework.tex
The findings from our study revealed several avenues for future work, which we propose in this section.

\subsection{Toward \textit{optimally} communicating each spatial awareness type.}

Our findings showed that communicating some aspect of spatial awareness well is not simply about doing so to the maximum extent possible. For example, when it comes to conveying the presence of items (Type~4), the simple audio menu facilitated it perfectly (receiving perfect Likert scores), but many participants disliked how it listed every item within the room they were currently in. They thought that the menu communicated too much information --- enough to affect how much fun they had playing the game.

These findings indicate that --- particularly within video games --- communicating a specific type of spatial awareness optimally does not necessarily mean communicating it at the maximum possible level. Future work should address what "optimal" really means in terms of communicating each type of spatial awareness. For example, in the case of item presence information (Type 4): What is the proper level of item presence that should be communicated to the player, and what factors --- such as game objectives --- may influence the level of item presence a tool should communicate? Similar questions can be extended to the other types as well.

\subsection{Toward purpose-built hardware.}
\label{sec:hardwarerw}

Participants revealed that they disliked juggling multiple pieces of hardware while using the smartphone map. Future touch-based SATs should reduce the number of devices required. One possibility involves using touchpads found on game controllers such as the DualShock 4~\cite{Dualshock4}, DualSense~\cite{Dualsense}, and Steam Controller~\cite{SteamController}. Hybrid touchscreen controller devices, such as the Nintendo Switch~\cite{NintendoSwitch} and the Steam Deck~\cite{SteamDeck}, are also promising alternatives.

\subsection{Applications for physical world navigation.}

In Section \ref{sec:affresults}, we addressed RQ1 by reporting participants' preferences for the six types of spatial awareness within a video game context. Future work could explore VIPs' preferences within the \textit{physical} world and see how they differ from their preferences within video games. We found that some of our results resemble prior work in the physical world: Participants generally agreed that position \& orientation was extremely important --- in our study, they collectively saw it as the \textit{most} important type of spatial awareness. In a similar vein, much physical world work for both visually impaired and sighted people has found position \& orientation awareness to be very important~\cite{Klatzky1998, Epstein2017, Giudice2020, GiudiceLegge2008, Kacorri2016}. Participants also generally found knowledge of item presence, item arrangement, and adjacent areas to be relatively important as well --- reflecting prior work that has echoed the importance of inter-object and inter-area relationships in promoting spatial awareness~\cite{GiudiceLegge2008, Hill1993, RowellUngar2005, Yatani2012}.

We were surprised, however, when we found that participants did not find scale and shape awareness to be very important within video games. This differs from much prior work from physical world contexts --- especially in the realm of tactile maps and echolocation --- that has found general overviews of spaces, including information such as scale and shape, to be crucial for spatial awareness~\cite{RowellUngar2005, HolmesArditi1998}.

An interesting direction for future work may involve repeating the study presented in this paper, but in the physical world, to enable a direct comparison. The physical world presents its own challenges and circumstances. SATs' accuracy within physical environments and VIPs' physical safety considerations~\cite{Banovic2013}, for example, could influence how important VIPs find the various types of spatial awareness and even how they wayfind and explore using the tools. 

A direct comparison can help the community establish a hierarchy from the spatial information that we know is important to VIPs. It can also help the community establish formal principles for prioritizing the display of different types of information during physical world navigation.

%% file: sec09-limitations.tex
As with many studies that involve people with visual impairments, we had a low number of participants. The preferences for spatial awareness that we found are based on the perspective of our nine participants and may differ for other VIPs. Although the four SATs we implemented covered a broad range of design possibilities, there may be other designs that we did not consider that could reveal more insights into what VIPs value in a spatial awareness tool for virtual worlds. Furthermore, while we are grateful to the nine VIPs who participated in our study, we regret not being able to recruit a more diverse group of participants. Finally, our work focused on 3D adventure video games with large worlds that players can traverse, and our testbed did not feature any moving objects. As such, additional work is needed to investigate SATs that may assist VIPs within other types of video games, especially those that feature moving objects (for example, enemies and projectiles). 

%% file: sec10-conclusion.tex
In this work, we explore the merits and limitations of existing approaches to facilitating spatial awareness for VIPs within video game worlds in order to allow accessible game designers to have a better understanding of which spatial awareness tools are best to include in games to make them accessible to VIPs. Through a user study, we investigated four leading approaches to facilitating spatial awareness for VIPs in an effort to understand what aspects of spatial awareness VIPs find important within games (RQ1), and to determine how well today's differing SAT approaches facilitate the various aspects of spatial awareness (RQ2).  

Regarding the first question, we found that participants considered position and orientation to be the most important aspect of spatial awareness, and that scale and shape are the least important. Regarding the second question, participants found the directional scanner to communicate the arrangement of items very well, the simple audio menu to communicate the presence of items very well, the smartphone map to communicate the shape of areas very well, and the whole-room shockwave to communicate the scale of the area well. Our findings also revealed deficiencies in current SAT approaches, including that some tools tend to provide too much information and that no tool excels at communicating position and orientation information --- despite it being the most important type of spatial awareness to participants.

We hope that better understanding VIPs' preferences for spatial awareness as well as how today's SATs work can open up access to more mainstream 3D video games, granting VIPs the same gaming experiences that sighted players are so often afforded.